\documentclass[
reprint,
superscriptaddress,
showpacs,showkeys,preprintnumbers,
amsmath,amssymb,
aps,
pra,
]{revtex4-1}

\usepackage{graphicx}
\usepackage{bm}
\usepackage{hyperref}
\usepackage{enumerate}
\usepackage{titlesec}
\usepackage{dsfont}

\newcommand{\dr}{\rangle\!\rangle}
\newcommand{\dl}{\langle\!\langle}
\newcommand{\ea}[1]{\begin{align}#1\end{align}}
\newcommand{\mcl}[1]{\mathcal{#1}}
\newcommand{\mbf}[1]{\mathbf{#1}}
\newcommand{\qcn}{\mcl{C}}
\newcommand{\qst}{\mcl{Q}}
\DeclareMathOperator{\Tr}{Tr}

\usepackage{etoolbox}
\apptocmd{\sloppy}{\hbadness 10000\relax}{}{}

\begin{document}

\title{Quantum Conical Designs}

\author{Matthew A.\ Graydon}
\email{mgraydon[at]perimeterinstitute.ca}
\affiliation{Perimeter Institute for Theoretical Physics, 31 Caroline St.\ N., Waterloo, Ontario N2L 2Y5, Canada}
\affiliation{Institute for Quantum Computing, University of Waterloo, Waterloo, Ontario N2L 3G1, Canada}
\affiliation{Department of Physics $\&$ Astronomy, University of Waterloo, Waterloo, Ontario N2L 3G1, Canada}
\author{D.\ M.\ Appleby}
\email{marcus[at]physics.usyd.edu}
\affiliation{Centre for Engineered Quantum Systems, School of Physics, The University of Sydney, Sydney, NSW 2006, Australia}
\date{\today}
\begin{abstract}
Complex projective $t$-designs, particularly \textsc{sic}s and full sets of \textsc{mub}s, play an important role in quantum information. We introduce a generalization which we call conical $t$-designs.  They include arbitrary rank symmetric informationally complete measurements (\textsc{sim}s) and full sets of arbitrary rank mutually unbiased measurements (\textsc{mum}s). They are deeply implicated in the description of entanglement (as we show in a subsequent paper). Viewed in one way a conical 2-design is a symmetric decomposition of a separable Werner state (up to a normalization factor). Viewed in another way it is a certain kind of polytope in the Bloch body. In the Bloch body picture \textsc{sim}s and full sets of  \textsc{mum}s form highly symmetric polytopes (a single regular simplex in the one case; the convex hull of a set of orthogonal regular simplices in the other).  We give the necessary and sufficient conditions for an arbitrary polytope to be what we call a homogeneous conical $2$-design.  This suggests a way to search  for new kinds of \emph{projective} $2$-design.
\end{abstract}

\maketitle
\section{INTRODUCTION}
Complex projective $t$-designs \cite{Neumaier81, Hoggar82, Zauner99, Scott06} play an important role in quantum information (with the vast majority of current applications being for the case $t=2$). The best known examples are  full sets of mutually unbiased bases~\cite{Schwinger:1960,Ivanovic81,Wootters89,Durt:2010} (\textsc{mub}s), and the families of projectors defining rank 1 symmetric informationally complete measurements~\cite{Zauner99,Renes04A,Scott10} (\textsc{sic}s).  However, these are not the only examples; in particular, it is known that such designs exist in every finite dimension~\cite{Seymour:1984,Hayashi:2005}.  Among other things they have applications to quantum tomography~\cite{Wootters89,Hayashi:2005,Scott06,Durt:2010,Adamson:2010,FernandezPerez:2011,Zhu2012,Tabia:2012},  cryptography~\cite{Bennett84,Bruss:1998,Fuchs:2003,Renes:2005,Durt:2010,Mafu:2013}, dense-coding~\cite{Bennett92,Durt:2010}, teleportation~\cite{Bennett93,Durt:2010}, entanglement detection~\cite{Spengler12,Rastegin:2013,Chen:2015,Graydon:2015a}, quantum communication~\cite{DallArno:2014a,DallArno:2015,Slomczynski:2014,Szymusiak:2014} and cloning~\cite{Gisin97,Cerf:2002,Scott06,Durt:2010} (references given being representative only).  They are important in quantum foundations, \textsc{sic}s being a mathematical cornerstone of QBism~\cite{Fuchs10,Fuchs13}.  They also have important applications in \emph{classical} information, in particular to compressed sensing~\cite{Balan:2009,Gross:2015}.  

A complex projective $t$-design consists of rank 1 projectors.  In this paper we introduce a generalization which has many of the same properties, but in which the projectors are replaced by arbitrary  positive semi-definite operators.  The set of such operators is the cone generated by quantum state space  (\emph{i.e.}\ the set of all operators of the form $\lambda \rho$, with $\lambda\ge 0$ and $\rho$ a density matrix).  We accordingly refer to the structures we introduce as conical $t$-designs.  Although our definition is valid for arbitrary $t$ we will focus in this paper on the case $t=2$, reserving the case $t>2$ for later work.   We will therefore use ``conical design'' as shorthand for ``conical $2$-design'', and ``projective design'' as shorthand for ``complex projective $2$-design''.  

One  example of a conical design is the set of  projectors forming a weighted  projective design~\cite{Levenshtein:1998,Scott06,Roy:2007,McConnell:2008} (when appropriately re-scaled).  Other examples are arbitrary rank symmetric informationally complete measurements (\textsc{sim}s) introduced in ref.~\cite{Appleby07} (under the different acronym \textsc{si-povm}s) and full sets of arbitrary rank mutually unbiased measurements (\textsc{mum}s) introduced in ref.~\cite{Kalev14b}.  We thus provide an affirmative answer to the question posed by Dall'Arno~\cite{DallArno14}, whether \textsc{sim}s and \textsc{mum}s are particular instances of a more general class of objects.    

Conical designs are interesting in their own right.  In particular, we will show in a subsequent publication~\cite{Graydon:2015a} that they are deeply implicated in the description of entanglement. We show in  Section~\ref{secWerner}  that they provide simple decompositions of some separable Werner~\cite{Werner:1989} and isotropic~\cite{Horodecki:1999} states.  Actually it is easy to see that conical designs provide simple expansions of all Werner and isotropic states (including the entangled ones), although, it remains to be seen whether such expansions can be put to any interesting use.  Another reason for being interested in them is their potential relevance to  projective designs.  The problem of constructing projective designs is  difficult.  The \textsc{mub} and \textsc{sic} existence problems are still open, and in many dimensions the only known examples have a cardinality which grows extremely fast with dimension~\cite{Seymour:1984,Hayashi:2005,Ambainis:2007}.  As we discuss in Section~\ref{BlochHom}, embedding the class of  projective designs in a larger class of structures having nice mathematical properties casts some new and possibly fruitful light on the problem of constructing new projective designs.

The structure of the paper is as follows.  We begin with two introductory sections.  The generalized Bloch representation~\cite{Harriman78,Mahler95,Siennicki01,Kimura03,Schirmer04,Kimura05,Dietz06,Appleby07} of quantum state space is central to our analysis.  In  Section~\ref{secBloch} we accordingly summarize the essential features of this representation.  In Section~\ref{secMumSim} we go on to describe \textsc{sim}s and \textsc{mum}s in terms of the Bloch picture.  The existence of \textsc{sim}s in every finite dimension was originally proved in ref.~\cite{Appleby07} using a simple geometrical argument.  It was subsequently re-proved in ref.~\cite{Kalev14a}  using a more complicated algebraic argument.  In ref.~\cite{Kalev14b}   similar algebraic methods were used to prove the existence of \textsc{mum}s in every finite dimension.  Since our analysis of conical designs relies on the geometrical approach we begin our discussion by giving a unified geometrical description of \textsc{sim}s and \textsc{mum}s.  The main part of the paper starts in Section~\ref{conical2Designs} where we define conical designs, and derive some basic properties.   The class of all conical designs is  large, and to make progress one needs to focus on some special cases.  One important special case is the class of weighted projective designs; these, however, have already been extensively discussed in the literature~\cite{Levenshtein:1998,Scott06,Roy:2007,McConnell:2008}.  In Section~\ref{BlochHom} we therefore examine another special case, which we call homogeneous conical designs.  Projective designs are examples of homogeneous designs, as are \textsc{sim}s and \textsc{mum}s.  In the Bloch body picture \textsc{sim}s and full sets of \textsc{mum}s form simple, highly symmetric polytopes (a single regular simplex in the case of \textsc{sim}s; the convex hull of a set of orthogonal regular simplices in the case of \textsc{mum}s).  We fully characterize the polytope corresponding to an arbitrary homogeneous design.  We further show that the problem of constructing a homogeneous  \emph{two}-design on a complex vector space reduces to the problem of constructing a  \emph{one}-design in a higher dimensional real vector space.  We discuss how these geometrical results suggest a way to systematically search for new examples of projective designs.  
In Section~\ref{secWerner} we show that conical designs provide simple decompositions of some separable Werner and isotropic states.  In this connection let us note that, although it is the entangled Werner states which are the most interesting, the problem of decomposing the separable states is not straightforward and has attracted some notice in the literature~\cite{Azuma:2006,Unanyan:2007,Tsai:2013}. In Section~\ref{secConc} we discuss some possible avenues of further research.

\section{THE BLOCH BODY}
\label{secBloch}
Throughout $\mathcal{H}$ will be a fixed $d$-dimensional complex Hilbert space.  We define
\begin{itemize}
\item  $\mathcal{L}$ to be the complex vector space consisting of all linear operators on $\mathcal{H}$,
\item  $\mathcal{L}_{sa}$ to be the real vector space consisting of all self-adjoint operators on $\mathcal{H}$,
\item $\mathcal{L}_{sa,0}$ to be the subspace of $\mathcal{L}_{sa}$ consisting of all trace zero operators,
\item $\mathcal{L}_{sa,1}$ to be the hyperplane in $\mathcal{L}_{sa}$ consisting of all trace one operators,
\item  $\qcn$ to be the  cone consisting of all positive semi-definite operators in $\mathcal{L}_{sa}$,
\item $\qst$ to be quantum state space---\emph{i.e.} the intersection $\qcn \cap\;\mathcal{L}_{sa,1}$.
\end{itemize} 
We equip $\mathcal{L}_{sa}$ with the Hilbert-Schmidt inner product
\ea{
\dl A_1 | A_2 \dr &= \Tr(A_1 A_2).
}
and the associated norm $\| A\| = \sqrt{\dl A | A \dr }$.

For a qubit $\qst$ can be identified with the Bloch ball.  For $d>2$ the geometry of $\qst$ is much more intricate than  a simple ball~\cite{Bengtsson08}.  Nevertheless, it is possible to construct a generalized Bloch representation, which preserves some essential features of the qubit case.  We will follow the coordinate-free approach of ref.~\cite{Appleby07} (also see refs.~\cite{Harriman78,Mahler95,Siennicki01,Kimura03,Schirmer04,Kimura05,Dietz06}).  

The key is to realize that  any $\rho\in \qst$ can be written in the form
\begin{equation}
\rho = \frac{1}{d} \left(I + B\right)
\end{equation}
for some $B \in \mathcal{L}_{sa,0}$.  In the $2$-dimensional case it is customary to write  $B= \mbf{n}\cdot \boldsymbol{\sigma}$ (where $\boldsymbol{\sigma}$ is the vector formed from the Pauli matrices) and to refer to $\mbf{n}$ as the Bloch vector corresponding to $\rho$.  In the general case it is more convenient to adopt a coordinate-free point of view, and to regard $B$ itself as the Bloch vector corresponding to $\rho$.  The Bloch body is then the set of all such  vectors:
\ea{
\mcl{B} &= \biggl\{ B \in \mcl{L}_{sa,0} \colon \frac{1}{d} \left(I+ B\right) \in \qst \biggr\}
}
The analogy with the 2-dimensional case is clearest if we introduce the scaled inner product and norm
\ea{
\dl B_1 | B_2 \dr_{\mcl{B}} &= \frac{1}{d(d-1)} \dl B_1 | B_2 \dr\text{,}
\\
\| B \|_{\mcl{B}} &= \frac{1}{\sqrt{d(d-1)}} \|B \|\text{.}
}
We then define the in- and out-balls
\ea{
\mathcal{B}_{\rm{in}}&=\left\{B\in\mathcal{L}_{sa,0}\colon \|B\|_{\mcl{B}} \leq \frac{1}{d-1} \right\}\text{,}
\label{inBall}
\\
\mathcal{B}_{\rm{out}}&=\left\{B\in\mathcal{L}_{sa,0} \colon \|B\|_{\mcl{B}} \leq 1\right\}\text{.}
}
We define $\mcl{S}_{\rm{in}}$ (respectively,  $\mcl{S}_{\rm{out}}$) to be the surface of $\mcl{B}_{\rm{in}}$ (respectively, $\mcl{B}_{\rm{out}}$).  One finds~\cite{Appleby07}
\ea{
\mcl{B}_{\rm{in}} \subseteq \mcl{B} \subseteq \mcl{B}_{\rm{out}}\text{.}
}
Moreover $\mcl{B}_{\rm{in}}$ (respectively, $\mcl{B}_{\rm{out}}$) is the largest (respectively, smallest) ball centred on the origin and contained in $\mcl{B}$ (respectively, containing $\mcl{B}$).  The manifold of pure states is the intersection $\mcl{B}\cap \mcl{S}_{\rm{out}}$.  If $d=2$ then $\mcl{B}_{\rm{in}} = \mcl{B}_{\rm{out}} = \mcl{B}$ and we recover the usual Bloch-ball description.

\section{MUBS AND SICS GENERALIZED}
\label{secMumSim}
The Bloch body picture gives a particularly intuitive way of thinking about \textsc{mub}s and \textsc{sic}s.  A full set of \textsc{mub}s is a family of $d(d+1)$ rank-1 projectors $\Pi_{b,j}$ (where $b=1,\dots, d+1$, $j=1, \dots, d$) such that.
\ea{
\dl \Pi_{b,j} |\Pi_{b',j'} \dr &= 
\begin{cases}
\delta_{j,j'} \qquad & b  = b' \text{,}
\\
\frac{1}{d} \qquad & b \neq b' \text{.}
\end{cases}
}
If $B_{b,j}$ are the corresponding Bloch vectors we have
\ea{
\dl B_{b,j} | B_{b',j'} \dr_{\mcl{B}} &=
\begin{cases}
\frac{d\delta_{j,j'}-1}{d-1} \qquad & b=b' \text{,}
\\
0 \qquad & b\neq b' \text{.}
\end{cases}
\label{eq:MUBBBdyDef}
}
The  vectors thus form $d+1$ orthogonal $d-1$ dimensional regular simplices with vertices in  $\mcl{B}\cap \mcl{S}_{\rm{out}}$.  

A \textsc{sic} is a  \textsc{povm} consisting of $d^2$ effects $E_j = (1/d) \Pi_j$ where the $\Pi_j$ are rank-$1$ projectors satisfying
\ea{
\dl \Pi_j | \Pi_{j'}\dr &= \frac{d\delta_{j,j'}+1}{d+1}\text{.}
}
If $B_j$ are the corresponding Bloch vectors we have
\ea{
\dl B_j | B_{j'} \dr_{\mcl{B}} &= \frac{d^2\delta_{j,j'}-1}{d^2-1}\text{.}
\label{eq:SICBBdyDef}
}
The  vectors thus form a single $d^2-1$ dimensional regular simplex with vertices in $\mcl{B}\cap \mcl{S}_{\rm{out}}$. 

The  existence problems for \textsc{mub}s and \textsc{sic}s are still open, notwithstanding the enormous amount of theoretical work which has been devoted to them.  Full sets of \textsc{mub}s have been shown to exist in every prime power dimension~\cite{Schwinger:1960,Ivanovic81,Wootters89}, but not in any other dimension.  Moreover there is much  evidence~\cite{Grassl04,Butterley07,Brierley08} supporting the conjecture~\cite{Zauner99} that a full set of \textsc{mub}s does not exist for $d=6$.  Turning to \textsc{sic}s, these have been constructed numerically~\cite{Renes04A,Scott10,ScottUP} for every $d \le 121$ and exact solutions~\cite{Zauner99,Grassl04,Appleby05,Grassl05,Grassl08,Scott10} have been constructed for $d=2-16$, $19, 24, 28, 35, 48$.  This encourages the conjecture that \textsc{sic}s  exist in every finite dimension.  

The Bloch body picture provides us with a simple geometrical explanation of why the \textsc{mub} and \textsc{sic} existence problems are so hard.  It is easy to construct vectors satisfying  Eqs.~(\ref{eq:MUBBBdyDef}) and (\ref{eq:SICBBdyDef})  if one only requires that they lie in the $d^2-2$ dimensional manifold $\mcl{S}_{\rm{out}}$.   What is hard (if $d>2$) is then to rotate the vectors so that they all lie in the measure zero, $2d-2$ dimensional submanifold $\mcl{B}\cap \mcl{S}_{\rm{out}}$.   

The triviality of the problem of inscribing a regular simplex into a sphere  motivated one of us~\cite{Appleby07} to introduce the concept of a \textsc{sim}, or arbitrary-rank symmetric informationally complete measurement (what in ref.~\cite{Appleby07} was called an \textsc{si}-\textsc{povm}).  Suppose that in the definition of a \textsc{sic} one drops the requirement that the effects are rank-$1$, so that one is only looking for a \textsc{povm} which is
\begin{enumerate}
\item Informationally complete.
\item Symmetric in the sense that the effects satisfy $\dl E_j | E_{j'}\dr = \alpha \delta_{j,j'} + \beta$ for some $\alpha$, $\beta$.
\end{enumerate}
In terms of the Bloch body description this means~\cite{Appleby07} that, instead of looking for Bloch vectors satisfying Eq.~(\ref{eq:SICBBdyDef}), one is only demanding
\ea{
\dl B_j | B_{j'} \dr_{\mcl{B}} &= \frac{\kappa^2(d^2\delta_{j,j'}-1)}{d^2-1} \text{.}
\label{eq:SIMBBdyDefB}
} 
for some $\kappa$ in the interval $(0,1]$.  In other words one is still looking for a regular simplex in $\mcl{B}$; however, one  no longer insists that the vertices lie on the out-sphere.  We refer to such a structure as a \textsc{sim}, and to $\kappa$ as the contraction parameter.  As $\kappa$ is reduced the area of the intersection of the sphere of radius $\kappa$ with $\mcl{B}$ becomes larger, and so the problem of finding a \textsc{sim} becomes easier.  Moreover, the fact that $\mcl{B}_{\rm{in}}\subseteq \mcl{B}$ means that the problem becomes trivial once $\kappa \le 1/(d-1)$.  

In ref.~\cite{Appleby07}, in addition to the above simple, geometrical argument to show  \textsc{sim}s exist for all $d$ and all   $\kappa \le 1/(d-1)$, it was also shown that for $d$  odd there exist \textsc{sim}s with $\kappa = 1/\sqrt{d+1}$ (termed Wigner \textsc{povm}s on account of their intimate relation to the Wigner function). Gour and Kalev~\cite{Kalev14a} subsequently constructed \textsc{sim}s using a  more complicated algebraic method, involving the generalized Gell-Mann matrices.   In odd dimension greater than 3 their \textsc{sim}s have a  contraction parameter  significantly less than $1/\sqrt{d+1}$ and are therefore not an improvement on the Wigner \textsc{povm}.  In even dimension, on the other hand, their \textsc{sim}s are an improvement  on the ones constructed in ref.~\cite{Appleby07}.  For further discussion of \textsc{sim}s and their applications see \cite{Rastegin14,Zhu:2014a,Chen:2015}.

For the sake of completeness let us note that if $B_j$ is a set of Bloch vectors satisfying Eq.~(\ref{eq:SIMBBdyDefB}) then the corresponding \textsc{sim} consists of the $d^2$ effects
\ea{
E_j &= \frac{1}{d^2} \left( I + B_j \right)\text{.}
}
satisfying
\ea{
\dl E_j | E_{j'}\dr &= \frac{d^2\kappa^2 \delta_{j,j'} +d+1-\kappa^2}{d^3(d+1)}\text{.}
\label{simDef}
}

A  similar approach can be taken with \textsc{mub}s.  If we relax the requirement that the Bloch vectors lie on the out-sphere then Eq.~(\ref{eq:MUBBBdyDef}) becomes
\ea{
\dl B_{b,j} | B_{b',j'} \dr_{\mcl{B}} &=
\begin{cases}
\frac{\kappa^2(d\delta_{j,j'}-1)}{d-1} \qquad & b=b' 
\\
0 \qquad & b\neq b'
\end{cases}
\label{eq:MUMBBdyDef}
}
with $\kappa \in (0,1]$.  Given a solution to these equations the operators
\ea{
E_{b,j} &= \frac{1}{d}\left(I + B_{b,j}\right)
}
form a full set of \textsc{mum}s~\cite{Kalev14b} (mutually unbiased measurements).  If $\kappa < 1$ the $E_{b,j}$ are not rank-1 projectors.  However, they  still form a \textsc{povm} for each fixed $b$.  Moreover, the \textsc{povm}s are unbiased in the sense that, just as for a full set of \textsc{mub}s, $\dl E_{b,j} | E_{b',j'}\dr =1/d$ for $b \neq b'$:
\ea{
\dl E_{b,j} | E_{b',j'}\dr &=
\begin{cases}
 \kappa^2 \delta_{j,j'} + \frac{1-\kappa^2}{d} \qquad & b = b' \text{,}
 \\
 \frac{1}{d} \qquad & b \neq b' \text{.}
 \end{cases}
\label{mumDef}
}

As with \textsc{sim}s one immediately sees, from the basic geometrical properties of the Bloch body,  that full sets of \textsc{mum}s  exist for all $\kappa \le 1/(d-1)$.  In ref.~\cite{Kalev14b}, where these structures were first introduced, it was shown that full sets of \textsc{mum}s can in fact be constructed with
\ea{
\kappa &= \sqrt{\frac{2}{d(d-1)}}
}
(note that the definition of $\kappa$ in ref.~\cite{Kalev14b} is different from the one adopted here). For the application of \textsc{mum}s to entanglement detection see ref.~\cite{Chen:2014}.

\section{ENTER  CONICAL DESIGNS}
\label{conical2Designs}
A full set of \textsc{mub}s,  and the family of projectors defining a \textsc{sic}, are  examples of projective designs (recall that we are using ``projective design'' as a shorthand for ``complex projective 2-design'').  In this section we will introduce a more general kind of design having \textsc{mum}s and \textsc{sim}s as special cases.  

A projective design is a non-empty family of rank-$1$ projectors $\Pi_j$ such that $\sum_j \Pi_j \otimes \Pi_j$ commutes with $U\otimes U$ for every unitary $U$.  It is natural to ask what can be said of an arbitrary family of operators $A_j \in \qcn$ having this property.  Theorem 1 answers that question.  

Before stating the theorem it will be convenient to introduce some notation. We define
\begin{itemize}
\item  $\Pi_{\rm{sym}}$ and $\Pi_{\rm{asym}}$ to be, respectively, the projectors onto the symmetric and antisymmetric subspaces of $\mcl{H}\otimes \mcl{H}$,
\item $W$ to be the unitary swap operator on $\mcl{H}\otimes \mcl{H}$ which takes $|\psi\rangle \otimes |\phi\rangle$ to $|\phi\rangle \otimes |\psi\rangle$, 
\item    $\mbf{I}$ to be the identity superoperator on $\mcl{L}$.
\end{itemize}   
Relative to some fixed \textsc{onb} $|e_j\rangle$  we also define
\begin{itemize}
\item  $|\Phi_{+}\rangle$ to be the maximally entangled state $(1/\sqrt{d})\sum_j |e_j\rangle \otimes |e_j\rangle$,
\item $A^{*}$ to be, for given $A\in \mcl{L}$,   the operator $\sum_{j} \langle e_j | A | e_k\rangle^{*} |e_j\rangle\langle e_k |$, 
\item $\mbf{T}$ to be the transpose superoperator which acts on $\mcl{L}$ according to $\mbf{T}(|e_j\rangle\langle e_k |) = |e_k\rangle \langle e_j|$.
\end{itemize}
 
\vspace{6pt}
\noindent\textbf{Theorem 1.} \label{Theorem1} Let $\{A_{1},\dots,A_{m}\}$ be a  family of operators in $\qcn$. Then the following \ statements are equivalent
\begin{enumerate}[(i)]
\item $\sum_{j=1}^{m}A_{j}\otimes A_{j}$ commutes with $U\otimes U$ for every unitary $U$.
\item For some $k_{s}\geq k_{a}\geq 0$
\ea{
\sum_{j=1}^{m}A_{j}\otimes A_{j}&=k_{s}\Pi_{\text{sym}}+k_{a}\Pi_{\text{asym}}\text{.}
\label{cDesDefCond2}
}
\label{thm1B}
\item For some $k_{+}\ge k_{-}\ge 0$
\ea{
\sum_{j=1}^{m}A^{\vphantom{*}}_{j}\otimes A^{*}_{j}&= k_{+} I + dk_{-} |\Phi_{+}\rangle \langle \Phi_{+}|\text{.}
\label{isoSte}
}
\label{thm1C}
\item For some $k_{+} \ge k_{-} \ge 0$
\ea{
\sum_{j=1}^m |A^{\vphantom{*}}_j \dr \dl A^{*}_j | &= k_{+} |I\dr \dl I | + k_{-} \boldsymbol{T} \text{.}
\label{approxTrans}
}
\label{thm1D}
\item For some $k_{+} \ge k_{-} \ge 0$
\ea{
\sum_{j=1}^m |A_j \dr \dl A_j | &= k_{+} |I\dr \dl I | + k_{-} \boldsymbol{I} \text{.}
\label{fromAFZ}
}
\label{thm1E}
\end{enumerate}
If these  equivalent conditions are satisfied, then the quantities $k_{\pm}$ in conditions~(\ref{thm1C})--(\ref{thm1E}) are the same, and are related to the quantities $k_s$, $k_a$ in condition~(\ref{thm1B}) by $k_{\pm} = (k_s \pm k_a)/2$.  The $A_{j}$ span $\mathcal{L}_{sa}$ if and only if $k_{s}>k_{a}$ (equivalently, $k_{-}> 0$).  

\vspace{6pt}
\noindent\textit{Proof.} To see that (\ref{thm1B}) $\iff$ (\ref{thm1C}) observe that Eq.~(\ref{cDesDefCond2}) can be written
\ea{
\sum_{j=1}^{m}A_{j}\otimes A_{j}&= k_{+} I  + k_{-} W
\label{cDesDefCond2Alt}
}
with $k_{\pm} = (k_s\pm k_a)/2$.  Taking the partial transpose on both sides we obtain  Eq.~(\ref{isoSte}).

Now let $J$ be the Choi-Jamio{\l}kowski isomorphism~\cite{Choi:1975,Jamiolkowski:1972} which takes a superoperator $\boldsymbol{\Lambda}$ to
 \ea{
 J(\boldsymbol{\Lambda}) = \frac{1}{d} \sum_{j,k} \boldsymbol{\Lambda}\bigl(|e_j\rangle\langle e_k|\bigr) \otimes |e_j\rangle \langle e_k |
 }
One easily verifies that
\ea{
J^{-1} (I) &= d | I\dr \dl I | \text{,}
\\
J^{-1} (W) &= d \mbf{T} \text{,}
\\
J^{-1} \bigl(|\Phi_{+}\rangle \langle \Phi_{+}|\bigr) &= \mbf{I} \text{,}
\\
\intertext{and, for all $A$, $B\in \mcl{L}_{sa}$,}
J^{-1} \bigl( A\otimes B\bigr) &= d |A\dr \dl B^{*}| \text{.}
}
Consequently, applying $J^{-1}$ to both sides of Eq.~(\ref{cDesDefCond2Alt}) gives Eq.~(\ref{approxTrans}) while applying it to both sides of Eq.~(\ref{isoSte}) gives Eq.~(\ref{fromAFZ}).   

We have shown that statements (ii), (iii), (iv) and (v) are equivalent.  The implication 
$(\rm{ii})\implies(\rm{i})$ is immediate.  So it only remains to  show  that $(\rm{i})\implies(\rm{ii})$. To see this observe that (i) implies~\cite{Itzykson:1966}
\begin{equation}
\sum_{j=1}^{m}A_{j}\otimes A_{j}=k_{s}\Pi_{\text{sym}}+k_{a}\Pi_{\text{asym}}\text{.}
\label{cDesDefCond2AltB}
\end{equation}
To see that $k_{s}$ and $k_{a}$ satisfy the stated inequalities, let $|\Psi\rangle$ be an arbitrary normalized element of the antisymmetric subspace of $\mathcal{H}\otimes\mathcal{H}$.  Then
\begin{equation}
k_{a}=\sum_{j=1}^{m}\langle\Psi|A_{j}\otimes A_{j}|\Psi\rangle\geq 0\text{.}
\end{equation}
Moreover partially transposing and applying $J^{-1}$ to Eq.~(\ref{cDesDefCond2AltB}) gives Eq.~(\ref{fromAFZ}) with $k_{\pm} = (k_s\pm k_a)/2$.  So $(k_s-k_a)/2 = \sum_{j} |\dl A_j | B\dr|^2$ for all normalized $B\in \mcl{L}_{sa,0}$.  It follows that $k_s\ge k_a$.  It also follows that if $k_s=k_a$ then the $A_j$ are not a spanning set for $\mcl{L}_{sa}$.  If, on the other hand, $k_s>k_a$ then it follows from Eq.~(\ref{fromAFZ}) that $\sum_j|\dl A_j | B\dr|^2 \ge k_{-}\|B\|^2>0$ for all non-zero $B \in \mcl{L}_{sa}$, implying that the $A_j$ are a spanning set. $\square$

\vspace{6pt}
It is easily seen that the theorem generalizes to arbitrary families of operators in $\mcl{L}$ (with the appropriate modification of the conditions on $k_s$ and $k_a$).  Lemma~1 in ref.~\cite{Appleby13} is a partial version of this more general result, in which the positivity requirement is relaxed, but not the requirement of self-adjointness, and in which only two of the five equivalent conditions are stated.

It will be seen that, up to normalization, the right-hand sides of Eqs.~(\ref{cDesDefCond2}) and~(\ref{isoSte})  are, respectively,  separable Werner states~\cite{Werner:1989}, and  separable isotropic states~\cite{Horodecki:1999}.  We return to this point in Section~\ref{secWerner}.  It will also be seen that if $k_{+}=k_{-}$ then, up to normalization, the right-hand side of Eq.~(\ref{approxTrans}) is the structural approximation to the transpose superoperator
\ea{
\tilde{\mbf{T}} &= \frac{1}{d+1}|I\dr \dl I| + \frac{1}{d+1}\mbf{T}
}
introduced by Horodecki~\cite{Horodecki:2003}.  As Kalev and Bae~\cite{Kalev:2013} have noted this means that, if $|\psi_1\rangle, \dots, |\psi_{d^2}\rangle$  are the normalized vectors defining a \textsc{sic}, then $\tilde{\mbf{T}}$ has the minimum cardinality Kraus decomposition
\ea{
\tilde{\mbf{T}} (A) &= \sum_j B^{\vphantom{\dagger}}_j A B^{\dagger}_j  & B^{\vphantom{\dagger}}_j  &= \frac{1}{\sqrt{d}} |\psi^{\vphantom{*}}_j\rangle \langle \psi^{*}_j |
}

\vspace{6pt} Theorem 1 suggests the following definition:

\vspace{6pt}
\noindent\textbf{Definition.} A conical $2$-design (or conical design for short) is a family of non-zero operators $A_1, \dots, A_m\in \qcn$ satisfying the five equivalent conditions (i)--(v) in Theorem $1$ with $k_s > k_a$ (equivalently, $k_{-}>0$).

\vspace{6pt}
The requirement that the $A_j$ all be non-zero is  not essential, and is made for convenience only.
We require that $k_s > k_a$  so as to ensure that the $A_j$ are a spanning set.  Note that if it were a projective design which was in question, so that the $A_j$ were rank-1 projectors, it would be enough to require that the set was non-empty in order to ensure that it was a spanning set.  But in the more general case that is no longer so. 

In the same way one can define a conical $t$-design for $t>2$ to be a set of non-zero operators $A_j \in \mcl{C}$ such that  $\sum_j A_j^{\otimes^t}$ commutes with every unitary of the form $U^{\otimes^t}$, and then use Schur-Weyl duality~\cite{Goodman:1998} to derive an analogue of Eq.~(\ref{cDesDefCond2}).  However, in this paper we will confine ourselves to the case $t=2$, deferring a consideration of the general case to a later publication.

The fact that the $A_j$ are a spanning set means that $m\ge d^2$.   Using Eq.~(\ref{fromAFZ}) one easily derives the following expansion formula for an arbitrary operator $L$ 
\ea{
L &= \frac{1}{k_{-}} \sum_j \left( \Tr(A_j L) - \frac{k_{+} \Tr(A_j) \Tr(L)}{dk_{+}+ k_{-}}\right) A_j
}
If $m=d^2$ the expansion is unique; otherwise not.

Obviously every projective design is a conical design.  More generally a conical design $A_j$ has $k_a =0$ if and only if the $A_j$ are all rank 1, so that when suitably re-scaled they form a weighted projective design.  In fact, taking the trace on both sides of Eqs.~(\ref{cDesDefCond2}) and~(\ref{fromAFZ}) gives
\ea{
\sum_j \left(\bigl(\Tr(A^{\vphantom{2}}_j)\bigr)^2 -\Tr(A_j^2) \right)& = d(d-1) k_a
}
So $k_a = 0$ if and only if  $\bigl(\Tr(A^{\vphantom{2}}_j)\bigr)^2 = \Tr(A_j^2)$ for all $j$, which in turn  is true if and only if the $A_j$ are all rank 1. 

It is easily seen that a full set of \textsc{mum}s is a conical design.  In fact, let $E_{b,j}$ be such a set.  If we define $N = \sum_{b,j} |E_{b,j}\dr \dl E_{b,j} |$ it follows from Eq.~(\ref{mumDef}) that
\ea{
N |E_{b,j}\dr &=  \frac{1}{d}(d+1-\kappa^2) |I\dr + \kappa^2 | E_{b,j}\dr \text{.}
}
Since the $E_{b,j}$ are a spanning set this means 
\ea{
N =  \frac{1}{d}(d+1-\kappa^2) |I\dr \dl I | + \kappa^2 \mathbf{I} \text{.}
}
The claim now follows from Theorem~1.

A similar argument shows that every \textsc{sim} is a conical design.  However, for \textsc{sim}s we have the stronger statement,  that a \textsc{povm} of cardinality $d^2$ is a conical design if and only if it is a \textsc{sim}.  To show that the condition is necessary as well as sufficient let $E_{1},\dots,E_{d^{2}}$ be a  conical design which is also a \textsc{povm}.   Eq.~\eqref{fromAFZ} implies
\begin{equation}
\sum_{j=1}^{d^{2}}\mathrm{Tr}(E_{j})E_{j}=(dk_{+} +k_{-}) I =\sum_{j=1}^{d^{2}}(dk_{+}+k_{-})E_{j}\text{,}
\end{equation}
Since the $E_j$ are a basis this means $\Tr(E_j) = dk_{+}  + k_{-}$.  The fact that $\Tr(E_j)$ is a constant means we must also have $\Tr(E_j) = 1/d$.  Taking account of the inequalities $k_{+} \ge k_{-} > 0$ we deduce
\ea{
k_{+} &= \frac{d+1-\kappa^2}{d^2(d+1)} & k_{-} &= \frac{\kappa^2}{d(d+1)}
}
for some $\kappa \in (0,1]$.  
By another application of Eq.~\eqref{fromAFZ}
\ea{
\sum_{k=1}^{d^{2}}\mathrm{Tr}(E_{j}E_k)E_k&=
\sum_{k=1}^{d^2} \left(\frac{k_{+}}{d} + k_{-} \delta_{j,k} \right) E_k \text{.}
}
Since the $E_j$ are a basis this means
\ea{
\Tr(E_j E_k) &= \frac{d^2\kappa^2\delta_{jk} + d+1 -\kappa^2}{d^3(d+1)} \text{.}
}
Comparing with Eq.~(\ref{simDef}) we see that the $E_j$ are a \textsc{sim}. 

It is not possible to prove an equally strong statement for \textsc{mum}s.    A full set of \textsc{mum}s, scaled by a factor of $1/(d+1)$, is a \textsc{povm} and a conical design.  However, there are other \textsc{povm}s of cardinality $d(d+1)$ which are conical designs.  For instance, if $E_j$ is a \textsc{sim}  then the \textsc{povm} with effects
\ea{
E'_j &= 
\begin{cases}
\frac{1}{2} E_j \qquad & 1 \le j \le d^2 \text{,}
\\
\frac{1}{2 d} I \qquad & d^2 < j \le d(d+1) \text{.}
\end{cases}
\label{MUMCounterExample}
}
is a conical design of cardinality $d(d+1)$.

At this stage it will be helpful to introduce some new notation.  Let $A_1, \dots, A_m$ be an arbitrary conical design, and let $t_j = \Tr(A_j)$.  Then $A_j$ has the Bloch representation
\ea{
A_j &= \frac{t_j}{d} (I + B_j) \text{.}
}
Define $\kappa_j = \|B_j \|_{\mcl{B}}$ and
\ea{
t &= \sqrt{\frac{1}{m} \sum_j t_j^2} & \kappa &= \sqrt{\frac{1}{mt^2} \sum_j t_j^2 \kappa_j^2} \text{.}
}
So $t$ is the rms trace, and $\kappa$ is the weighted rms Bloch vector norm.  Note that $\kappa \in (0,1]$, and that $\kappa =1$ if and only if the $A_j$ are all rank 1.  As with \textsc{sim}s and \textsc{mum}s we will refer to $\kappa$ as the contraction parameter.  Taking the trace on both sides of Eqs.~(\ref{cDesDefCond2}) and~(\ref{fromAFZ})  we find
\ea{
\frac{1}{2} d(d+1) k_s + \frac{1}{2} d(d-1) k_a &= \sum_j \bigl(\Tr(A_j)\bigr)^2 \text{,}
\\
\frac{1}{2} d(d+1) k_s - \frac{1}{2} d(d-1) k_a &= \sum_j \Tr(A_j^2) \text{,}
}
from which it follows
\ea{
k_s &= \frac{mt^2}{d^2} \left( 1+\frac{(d-1)\kappa^2}{d+1} \right) \text{,}
&
k_a &= \frac{mt^2(1-\kappa^2)}{d^2} \text{.}
\label{kskaTermskapt}
}
Taking a partial trace on both sides of Eq.~(\ref{cDesDefCond2}) we find
\ea{
\sum_j t_j A_j &= \frac{mt^2}{d} I \text{.}
}
It follows that the operators
\ea{
E_j &= \frac{d t_j}{mt^2} A_j
\label{conicalEffects}
}
constitute a \textsc{povm}.  In the case when the $A_j$ have constant trace (but not in general) this \textsc{povm} is also a conical design.  

Lastly, the Bloch vectors satisfy
\ea{
\sum_j t_j^2 B^{\vphantom{2}}_j &= 0 \text{,}
\label{genBvecProp1}
\\
\sum_j t_j^2 |B^{\vphantom{2}}_j\dr \dl B^{\vphantom{2}}_j | &= \frac{mdt^2\kappa^2}{d+1} \Pi_{\mcl{B}} \text{,}
\label{genBvecProp2}
}
where
\ea{
\Pi_{\mcl{B}} &= \boldsymbol{I} - \frac{1}{d} |I \dr\dl I | 
}
is the Bloch projector---\emph{i.e.}\ the projector onto the subspace $\mcl{L}_{sa,0}$. 

\section{BLOCH GEOMETRY IN THE HOMOGENEOUS CASE}
\label{BlochHom}
The class of all conical designs is large, and to make progress one needs to focus on special cases.  One important special case is the class of weighted projective designs, concerning which much is known~\cite{Levenshtein:1998,Scott06,Roy:2007,McConnell:2008}.
In this section we consider another special case.  Specifically, we consider  conical designs which are homogeneous in the sense that $\Tr(A_j)$ and $\Tr(A^2_j)$ are constant, so that $t_j=t$, $\kappa_j = \kappa$ for all $j$.  This class of designs includes \textsc{sim}s and full sets of \textsc{mum}s.  It also includes all projective designs.  Specifically the projective designs are precisely the homogeneous conical designs for which 
\ea{
t &=  \kappa = 1
}
or, equivalently,
\ea{
k_s &= \frac{2m}{d(d+1)} & k_a &= 0 \text{.}
}

In the remainder of this section we study the Bloch geometry of a homogeneous conical design.  We know that the Bloch vectors of \textsc{sim}s and full sets of \textsc{mum}s form polytopes having a simple geometrical description.  We would like to describe  the polytope corresponding to an arbitrary homogeneous conical design.   The geometry of the polytope is fully specified by the Gram matrix $G$ with matrix elements
\ea{
G_{jk} &= \dl B_j | B_k \dr 
}
We will therefore focus on the problem of characterizing this matrix.

\vspace{6pt}

\noindent\textbf{Theorem 2.} \label{Theorem2} Let $B_1, \dots, B_m$ be a set of  vectors in $\mcl{B}$. Then the following statements are equivalent
\begin{enumerate}[(i)]
\item The $B_j$ are the Bloch vectors of a homogeneous conical design.
\item Their Gram matrix is of the form
\ea{
G &= \lambda P
\label{homC2DprojDef}
}
where $\lambda$ is a positive constant and $P$ is a rank $d^2-1$ projector which is constant on the diagonal and such that $\sum_k P_{jk} = 0$ for all $j$.
\end{enumerate}
If these equivalent conditions are satisfied then $\lambda \le md/(d+1)$ and
$
P_{jk} \le (d^2-1)/m
$
for all $j$, $k$ with equality when $j=k$.  The associated conical designs  have contraction parameter 
\ea{
\kappa = \sqrt{\frac{\lambda(d+1)}{md}} \text{.}
}

\vspace{6pt}
\noindent \emph{Remark.} Notice that if the conditions of the theorem are satisfied then there are infinitely many  conical designs with Bloch vectors $B_j$ since the trace $t$ in $A_j = (t/d)(I+B_j)$ can take any positive value.

\vspace{6pt}
\noindent\emph{Proof.}   To show that $(\rm{i}) \implies (\rm{ii})$, suppose
\ea{
A_j &= \frac{t}{d} (I + B_j)
}
is a homogeneous conical design with contraction parameter $\kappa$.  It follows from Eq.~(\ref{genBvecProp2}) that
\ea{
\sum_j |B_j\dr \dl B_j | &= \lambda \Pi_{\mcl{B}} 
\label{homBvecProp2}
}
where $\lambda = md \kappa^2/(d+1)$.  So
\ea{
G^2 &=\lambda G \text{,}
}
implying that $P = (1/\lambda) G$
is a projection operator.  Taking the trace on both sides of Eq.~(\ref{homBvecProp2}) we find
\ea{
\Tr(P) &= \frac{1}{\lambda} \sum_{j} G_{jj} = d^2-1 \text{.}
}
So $P$ is rank $d^2-1$.  Moreover,
\ea{
P_{jj} &= \frac{1}{\lambda} \dl B_j | B_j \dr = \frac{d^2-1}{m} 
}
for all $j$.  So $P$ is constant on the diagonal.  Finally, it follows from Eq.~(\ref{genBvecProp1}) that $\sum_k B_{k}=0$, implying that $\sum_k P_{jk} = 0$ for all $j$.

To show that $(\rm{ii}) \implies (\rm{i})$, suppose the Gram matrix has the stated form.  Observe that the fact that the rank of the Gram matrix is $d^2-1$ means that the $B_j$ are a spanning set for $\mcl{L}_{sa,0}$. 
Let $N = \sum_j | B_j \dr \dl B_j |$.  Then
\ea{
\dl B_j | N | B_k \dr &= \lambda \dl B_j | B_k\dr
}
for all $j, k$.  Since the $B_j$ are a spanning set for $\mcl{L}_{sa,0}$, and  since $N |I\dr = 0$, this implies
\ea{
\sum_j |B_j \dr \dl B_j | &= \lambda \Pi_{\mcl{B}} \text{.}
}
The fact that $\sum_k P_{jk} = 0$ means $\sum_k B_k = 0$.  So if we define $A_j = (t/d) (I + B_j)$ for any fixed positive $t$ we will have
\ea{
\sum_j |A_j \dr \dl A_j | &= 
 \frac{t^2(md - \lambda)}{d^3}|I\dr \dl I | + \frac{t^2\lambda}{d^2} \text{.}
}
If we can show that $\lambda \le md/(d+1) $ it will follow that the $A_j$ are a conical design.  To see that this is the case observe that the fact that $P$ is constant on the diagonal means
\ea{
m P_{jj} &= \Tr(P) = d^2-1
}
for all $j$.  Consequently
\ea{
1 \ge \|B_j \|^2_{\mcl{B}} &= \frac{\lambda P_{jj}}{d(d-1)} = \frac{\lambda(d+1)}{md}
}
from which the claim follows.  We have incidentally shown that the design is homogeneous, with contraction parameter
\ea{
\kappa &= \sqrt{\frac{\lambda(d+1)}{md}} \text{.}
}

To prove the last part of theorem observe that the only statement not proved in the course of establishing the implication $(2)\implies (1)$ is the bound on the matrix elements of $P$.  This is an immediate consequence of the fact that $\| B_j \|^2_{\mcl{B}} = \lambda(d+1)/(md)$.  $\square$

\vspace{6pt}

Let $\mcl{P}_m$ be the set of all $m\times m$ rank $d^2-1$ projectors $P$ with the properties
\begin{align}
\text{1}.\hspace{0.1cm}& \sum_k P_{jk} = 0 \text{ for all } j\text{.} \label{PmCon1}\\
\text{2}.\hspace{0.1cm}& P_{jk} \le \frac{d^2-1}{m} \text{ for all } j,k \text{ with equality when } j=k.\label{PmCon2}
\end{align}
We have shown that \emph{some} projectors of this type are associated to homogeneous conical designs via Eq.~(\ref{homC2DprojDef}).  It remains to show that \emph{all} of them are, for every  $m\ge d^2$.

For given $m \ge d^2$ and  $P\in \mcl{P}_m$ let $S(P)$ be the set of all $m$-tuples of associated  vectors in $\mcl{B}$.  Thus $B = (B_1, \dots, B_m)\in S(P)$ if and only if 
\ea{
\dl B_j | B_k\dr &=\lambda P_{jk}
}
for some positive $\lambda$.  Also define, for each $B \in S(P)$,
\ea{
\kappa_B &= \|B_1\|_{\mcl{B}} = \dots = \|B_m\|_{\mcl{B}}
}
and let $K(P) = \{\kappa_{B} \colon B \in S(P)\}$. 
The convexity of the Bloch body means that if $B \in S(P)$  then so does $\eta B$ for all $\eta \in (0,1]$ (so  $S(P)$ is either empty or infinite).   It follows that if $\kappa \in K(P)$ then $(0,\kappa] \subseteq K(P)$.  So if 
\ea{
c_P &= 
\begin{cases}
\sup\bigl(K(P)\bigr) \qquad & \text{$K(P)$ non-empty}
\\
0 \qquad & \text{$K(P)$ empty}
\end{cases}
}
then $(0,c_P) \subseteq K(P) \subseteq (0,c_P]$.  We claim that in fact $K(P) = (0,c_p]$.  The claim is trivial if $c_P =0$, so we may assume without loss of generality that $c_P> 0$.   Choose a sequence $B_n \in S(P)$  such that $\kappa_{B_n} \uparrow c_P$.  Since $\mcl{B}^m$ is a  closed, bounded subset of a finite dimensional, real inner-product space it is compact~\cite{Mendelson:1962}.     We can therefore choose a convergent subsequence $B_{n_a}\to B \in \mcl{B}^m$.  We have
\ea{
\dl B_j | B_k\dr &=  \lim_{a\to \infty} \left( \frac{md \kappa_{B_{n_a}}^2}{d+1} \right) P_{jk} = \frac{md c_P^2}{d+1} P_{jk} \text{.}
}
So $B \in S(P)$ and $c_P = \kappa_B  \in K(P)$.  
 
It is known~\cite{Seymour:1984,Hayashi:2005} that projective designs exist in every dimension 
(although it should be noted~\cite{Ambainis:2007} that the cardinality of the projective designs constructed in these papers grows extremely fast with dimension).  
Since $c_P =1$ for the projector corresponding to a projective design this means that homogeneous conical designs exist in every dimension and for every  $\kappa \in (0,1]$.  

We are now in a position to prove the second main result of this section (which can be regarded as a generalization of the existence proofs for \textsc{sim}s and full sets of \textsc{mum}s).

\vspace{6pt}
\noindent\textbf{Theorem 3.} \label{Theorem3}  For all $m\ge d^2$ and $P\in \mcl{P}_m$
\ea{
c_P \ge \frac{1}{d-1} \text{.}
}
In particular $S(P)$ is non-empty.

\vspace{6pt}
\noindent\emph{Proof.} The fact that $P$ is a rank $d^2-1$ projector means we can choose $d^2-1$  orthonormal vectors $\vec{u}_a \in \mathbb{R}^m$ such that
\ea{
P_{jk} &= \sum_{a}^{d^2-1} u_{a,j}u_{a,k}
}
Let $D_1, \dots , D_{d^2-1}$ be an orthonormal basis for $\mcl{L}_{sa,0}$ and define $B = (B_1, \dots, B_m)$ by
\ea{
B_j & = \sqrt{\frac{md}{(d+1)(d-1)^2}} \sum_{a=1}^{d^2-1} u_{a,j} D_a
}
Then
\ea{
\dl B_j | B_k \dr &= \frac{md}{d+1)(d-1)^2} P_{jk}
}
In particular $\| B_j \|_{\mcl{B}} = 1/(d-1)$, implying that $B_j \in \mcl{B}_{\rm{in}}\subseteq \mcl{B}$.  So  $B \in S(P)$ and $1/(d-1) = \kappa_B \le c_P$. $\square$

\vspace{6pt}
In this paper we are mainly focusing on conical designs.  However, the result just established is potentially relevant to the problem of constructing projective designs. The projectors in $\mcl{P}_m$ which correspond to projective designs are precisely the ones for which $c_P = 1$.  This suggests the following program:
\begin{enumerate}
\item Classify the polytopes described by the projectors in $\mcl{P}_m$.
\item Identify those polytopes for which $c_P=1$.
\end{enumerate} 
This program is, of course, extremely ambitious as success would carry with it, as a minor corollary, solutions to the \textsc{mub} and \textsc{sic} existence problems.  However,  even some partial results  might be useful.  It might, for instance, be useful if one could exclude some of the projectors in $\mcl{P}_m$, as definitely not having $c_P=1$. One obvious way to do this  is to exploit the fact~\cite{Kimura05} that each vertex of the polytope corresponding to a projective design must be diametrically opposite a face which is tangential to $\mcl{S}_{\rm{in}}$.  Having narrowed down the set of candidates, one might then  investigate the remaining polytopes numerically, to see if any of them correspond to projective designs in low dimension.  In essence, this procedure---writing down a set of equations motivated by considerations of symmetry, and then looking for solutions in low dimension---was the way \textsc{sic}s were originally found~\cite{Zauner99,Renes04A}.  The same procedure might possibly be used to find other projective designs.

We conclude with a result which says  that the problem of constructing a homogeneous \emph{two}-design in a complex vector space reduces to the problem of constructing a \emph{one}-design in a higher-dimensional real vector space.

\vspace{6pt}
\noindent\textbf{Theorem 4.} \label{Theorem4}  Let $B_1, \dots, B_{m}$ be a set of Bloch vectors.  Then the following statements are equivalent
\begin{enumerate}[(i)]
\item The $B_j$ are the Bloch vectors of a homogeneous conical design.
\item The $B_j$ have the same norm and satisfy
\ea{
\sum_{j} B_j &= 0
\label{BvecsCentredProp}
\\
\sum_j |B_j \dr \dl B_j | &= \lambda \Pi_{\mcl{B}}
\label{onedesProp}
}
for some $\lambda>0$.
\end{enumerate}

\vspace{6pt}
\noindent\emph{Proof.}  The implication $(\rm{i})\implies(\rm{ii})$ is an immediate consequence of Eqs.~(\ref{genBvecProp1}) and~(\ref{genBvecProp2}).  To prove the converse let $B_j$ be a set of Bloch vectors having the stated properties and define
\ea{
P_{jk} &= \frac{1}{\lambda} \dl B_j | B_k \dr \text{.}
}
Then it follows from Eq.~(\ref{onedesProp}) that $P^2=P$, implying that $P$ is a projector.  Taking the trace on both sides of Eq.~(\ref{onedesProp}) we find
\ea{
\Tr(P) &= \frac{1}{\lambda} \sum_j \dl B_j | B_j \dr = \Tr(\Pi_{\mcl{B}}) = d^2-1 \text{.}
}
So $P$ is rank $d^2-1$.  The fact that the $B_j$ have the same norm means that $P$ is constant on the diagonal, while Eq.~(\ref{BvecsCentredProp}) implies that $\sum_k P_{jk} = 0$ for all $j$.  So it follows from Theorem $2$ that the $B_j$ are the Bloch vectors of a conical design. $\square$

\section{WERNER AND ISOTROPIC STATES}
\label{secWerner}
In Section~\ref{conical2Designs} we observed that, up to normalization, the right-hand sides of Eqs.~(\ref{cDesDefCond2}) and~(\ref{isoSte})  are, respectively,  separable Werner states~\cite{Werner:1989}, and  separable isotropic states~\cite{Horodecki:1999}. This   merits a little discussion.

A Werner state is one of the form
\ea{
\rho_W &= k_s \Pi_{\rm{sym}} + k_a \Pi_{\rm{asym}} 
\label{wernerDef}
}
with
\ea{ k_s &=\frac{2(1-p)}{d(d+1)} & k_a &= \frac{2p}{d(d-1)} 
}
for some $p\in [0,1]$.  The state is entangled if and only if $p \in (1/2,1]$. The entangled states are the ones of most interest since, in addition to Werner's original motivation, it can be shown~\cite{Horodecki:1999} that the existence of bound-entangled \textsc{npt} states is equivalent to the existence of bound-entangled Werner states.  The existence of the latter is still an  open question, but there are indications~\cite{DiVincenzo:2000a,Dur:2000} that the entanglement becomes bound as one approaches the cross-over point at $p=1/2$. As we remarked in the introduction conical designs can be used to provide simple decompositions of all Werner states, both separable and entangled (although it remains to be seen how interesting they are).  However, we will here confine ourselves to the point, which is already apparent from the definition, that they provide simple decompositions of some of the separable states.  In this connection let us observe that, although less interesting, the problem of decomposing a separable Werner state is not straightforward, and  has attracted some notice in the literature~\cite{Azuma:2006,Unanyan:2007,Tsai:2013}.  Conical designs cast  additional light on the problem.

Let us define a symmetric decomposition of a separable Werner state to be one of the form
\ea{
\rho_W &= \sum_{j=1}^m \lambda_j \rho_j \otimes \rho_j
\label{wernerDecompA}
}
where $\rho_j \in \mcl{Q}$, $\lambda_j \in (0,1]$ and $\sum_j \lambda_j = 1$.  We will say that the decomposition is homogeneous if $\lambda_j = 1/m$ for all $j$, and that it is pure if the $\rho_j$ are all pure.  It follows from Theorem 1 that $\rho_W$ does not have a symmetric decomposition if  $k_s < k_a$ or, equivalently, if $p > (d-1)/(2d)$.  If $p=(d-1)/(2d)$ then $\rho_W$ is  the maximally mixed state, so the existence of a symmetric decomposition is trivial.  If $p< (d-1)/(2d)$ then Eq.~(\ref{wernerDecompA}) is equivalent to the statement that the operators $A_j = \sqrt{\lambda_j} \rho_j$ are a conical design.  

It was shown in Section~\ref{BlochHom} that homogeneous conical designs exist for all $d$ and all $\kappa \in (0,1]$.   We conclude that a separable Werner state has a symmetric decomposition if and only if $0 \le p \le (d-1)/(2d)$.  Furthermore, if $p$ is in this interval the decomposition can always be chosen to be homogeneous.  Finally, it was shown in Section~\ref{conical2Designs} that a  conical design is rank 1 if and only if $k_a =0$ (in which case it is essentially the same thing as a weighted projective design~\cite{Levenshtein:1998,Scott06,Roy:2007,McConnell:2008}).  So $\rho_W$ has a pure symmetric decomposition if and only if $p=0$.  

We have thus shown that the interval $0\le p \le 1/2$ splits into two sub-intervals separated by the maximally mixed state at $p=(d-1)/2d$.  States in the sub-interval $0 \le p \le (d-1)/(2d)$ do have  symmetric decompositions;  states  in the sub-interval $ (d-1)/(2d) < p \le 1/2$ do not.
One motivation for studying separable Werner states is the hope that, by looking at the states immediately  below the cross-over at $p=1/2$, one may get some insight into the bound-entangled states conjectured to exist just above it.  From this point of view the most interesting feature of our discussion is the negative statement, that states immediately below the cross-over cannot be put into the simple form of Eq.~(\ref{wernerDecompA}).

In the case $p< (d-1)/(2d)$ we define an ideal decomposition to be one which  is symmetric, homogeneous and such that  $m$ achieves its minimum value of $d^2$.  A homogeneous conical design is a \textsc{povm} up to re-scaling, so we can use one of the results proved in Section~\ref{conical2Designs} to conclude that an ideal  decomposition must be of the form
\ea{
\rho_W &= \sum_{j=1}^{d^2} E_j \otimes E_j
}
where the $E_j$ constitute a \textsc{sim}.  In view of the discussion in Section~\ref{BlochHom} this gives us the following reformulation of the \textsc{sic}-existence problem:  A \textsc{sic} exists in dimension $d$ if and only if every Werner state with $0 \le p < (d-1)/2d$ has an ideal decomposition.

Conical designs can also be used to give simple decompositions of a subset of the isotropic states introduced in ref.~\cite{Horodecki:1999}. The states are defined by
\ea{
\rho_{I} & =  \frac{1-F}{d^2-1}  I +\frac{d^2F -1}{d^2-1} |\Phi_{+}\rangle \langle \Phi_{+} |
}
with $F\in [0,1]$ and $|\Phi_{+}\rangle$ the maximally entangled state defined at the beginning of Section~\ref{conical2Designs}. They are separable for $F\in [0,1/d]$ and entangled for $F \in (1/d, 1]$ (they are not, however, bound-entangled for any value of $F$).  
We define a symmetric decomposition of an isotropic state to be one of the form
\ea{
\rho_{I} &= \sum_{j=1}^m \lambda_j \rho_j \otimes \rho^{*}_j
\label{isotropicDecompA}
}
where $\rho_j \in \mcl{Q}$, $\lambda_j \in (0,1]$, and $\sum_j \lambda_j =1$. Symmetric decompositions of isotropic states are in bijective correspondence with symmetric decompositions of Werner states.  
In fact let 
\ea{
k_s \Pi_{\rm{sym}} + k_a \Pi_{\rm{asym}} = \sum_j \lambda_j \rho_j \otimes \rho_j
}
be a symmetric decomposition of a Werner state with $p$ in the interval $[0,(d-1)/(2d)]$.  Taking the partial transpose on both sides  gives
\ea{
k_{+} I + d k_{-} |\Phi_{+}\rangle \langle \Phi_{+} | &=  \sum_j \lambda_j \rho_j \otimes \rho^{*}_j
}
where $k_{\pm} = (k_s \pm k_a)/2$.  The fact that $0 \le p \le (d-1)/(2d)$ means $1/(d(d+1)) \le k_{+} \le 1/d^2$.  So we obtain in this way a symmetric decomposition of every isotropic state with $1/d^2 \le F \le 1/d$.  Reversing the argument it can be seen that, if one had a symmetric decomposition of an isotropic state with $0 \le F < 1/d^2$, then taking the partial transpose would give a symmetric decomposition of a Werner state with $(d-1)/(2d) < p \le 1/2$---which we have shown to be impossible.

Similarly to the Werner case, we see that the interval $0 \le F \le 1/d$ corresponding to the separable states splits into two sub-intervals, situated either side of the maximally mixed state at $F=1/d^2$.   States in the sub-interval $1/d^2 \le F \le 1/d$ do  have  symmetric decompositions;  states  in the sub-interval $ 0 \le F < 1/d^2$ do not.  The difference with the Werner case is that it is now the states \emph{with} a symmetric decomposition which lie next to the set of entangled states.

\section{CONCLUSION}
\label{secConc}

We introduced a new class of geometric structures in quantum theory, conical designs, which are natural generalizations of  projective designs.  We showed that \textsc{sim}s and \textsc{mum}s are special cases, as are weighted projective designs (up to re-scaling).  We began by establishing their basic properties.  In particular we gave five equivalent conditions for a set of positive semi-definite operators to be  a conical design (theorem~\hyperref[Theorem1]{1}).   We then turned to the special case of homogeneous conical designs, and analyzed their Bloch geometry.  In the Bloch body picture  \textsc{sim}s and full sets of \textsc{mum}s form simple, highly symmetric polytopes (a single regular simplex in the case of \textsc{sim}s; the convex hull of a set of orthogonal regular simplices in the case of \textsc{mum}s).  We showed that the same is true of an arbitrary homogeneous conical design.  Moreover, we derived necessary and sufficient conditions for a given polytope to be such a design (theorems~\hyperref[Theorem2]{2} and~\hyperref[Theorem3]{3}).   We also showed how the problem of constructing a homogeneous \emph{two}-design in a complex vector space reduces to the problem of constructing a spherical \emph{one}-design in a higher dimensional real vector space (theorem~\hyperref[Theorem4]{4}).  Finally, we showed that conical designs  provide simple decompositions of some separable Werner and isotropic states.

We show in a subsequent publication~\cite{Graydon:2015a} that conical designs are deeply implicated in the description of entanglement.  There are other questions which might be interesting to investigate.  Firstly, there  is our suggestion in Section~\hyperref[BlochHom]{V}, that the results there proved  could be used to search systematically for new \emph{projective} designs.  Secondly, all known examples of \textsc{sic}s and full sets of \textsc{mub}s  have important group covariance properties \cite{Dang:2015}.  One would like to know how far this holds true in the more general setting of homogeneous conical designs.  Thirdly, one would like to extend the analysis to conical $t$-designs with $t>2$ via Schur-Weyl duality~\cite{Goodman:1998}.  Fourthly, it is to be observed that the full class of conical 2-designs is itself a convex set.  It might be interesting to  explore the geometry of that set.  For instance, one might try to characterize the extreme points.  Finally,  it would be interesting to investigate conical designs in the larger context of general probabilistic theories~\cite{Barnum12}. 

\section{ACKNOWLEDGMENTS}
Research at Perimeter Institute is supported by the Government of Canada through Industry Canada and by the Province of Ontario through the Ministry of Research \& \mbox{Innovation}. MAG was supported by an NSERC Alexander Graham Bell Canada Graduate Scholarship. DMA was supported by the IARPA MQCO program, by the ARC via EQuS project number CE11001013, and by the US Army Research Office grant numbers W911NF-14-1-0098 and W911NF-14-1-0103.

\bibliography{cdRefARXv3}
\end{document}